\documentclass[%
 reprint,
 amsmath,amssymb,
 aps,
]{revtex4-2}

\usepackage{graphicx}% Include figure files
\usepackage{dcolumn}% Align table columns on decimal point
\usepackage{bm}% bold math
\usepackage{orcidlink}
\usepackage{hyperref}% add hypertext capabilities
\hypersetup{
    colorlinks=true,
    linkcolor=blue,
    urlcolor=black,
    citecolor=blue
    }
\begin{document}

\preprint{APS/123-QED}

\title{Second-Order Coherence Across the Brillouin Lasing Threshold}% Force line breaks with \\

\author{E.~A.~Cryer-Jenkins\,\orcidlink{0000-0003-2549-0280}\,$^{1,\dagger}$}
\author{G.~Enzian\,\orcidlink{0000-0002-2603-2874}\,$^{1,2,3,\dagger}$}
\author{L.~Freisem\,\orcidlink{0000-0002-9257-1243}\,$^{1,2}$} 
\author{N.~Moroney\,\orcidlink{0000-0002-9818-5351}\,$^{1}$}
\author{J.~J.~Price\,\orcidlink{0000-0001-9385-2883}\,$^{1,2}$}
\author{A.~\O.~Svela\,\orcidlink{0000-0002-3534-3324}\,$^{1,2}$}
\author{K.~D.~Major\,\orcidlink{0000-0002-3268-6946}\,$^{1}$}
\author{M.~R.~Vanner\,\orcidlink{0000-0001-9816-5994}\,$^{1,}$}
\email{www.qmeas.net (m.vanner@imperial.ac.uk)}

\address{$^1$QOLS, Blackett Laboratory, Imperial College London, London SW7 2BW, United Kingdom\\
$^2$Clarendon Laboratory, Department of Physics, University of Oxford, Oxford OX1 3PU, United Kingdom\\
$^3$Niels Bohr Institute, University of Copenhagen, Copenhagen 2100, Denmark\\
$^{\dagger}$These authors contributed equally and have been listed alphabetically.
}

\begin{abstract}
Brillouin--Mandelstam scattering is one of the most accessible nonlinear optical phenomena and has been widely studied since its theoretical discovery one hundred years ago. The scattering mechanism is a three-wave-mixing process between two optical fields and one acoustic field and has found a broad range of applications spanning microscopy to ultra-narrow-linewidth lasers. Building on the success of utilizing this nonlinearity at a classical level, a rich avenue is now being opened to explore Brillouin scattering within the paradigm of quantum optics. Here, we take a key step in this direction by employing quantum optical techniques yet to be utilized for Brillouin scattering to characterize the second-order coherence of Stokes scattering across the Brillouin lasing threshold. We use a silica microsphere resonator and single-photon counters to observe the expected transition from bunched statistics of thermal light below the lasing threshold to Poissonian statistics of coherent light above the threshold. Notably, at powers approaching the lasing threshold, we also observe super-thermal statistics, which arise due to instability and a ``flickering” in and out of lasing as the pump field is transiently depleted. The statistics observed across the transition, including the ``flickering'', are a result of the full nonlinear three-wave-mixing process and cannot be captured by a linearized model. These measurements are in good agreement with numerical solutions of the three-wave Langevin equations and are well demarcated by analytical expressions for the instability and the lasing thresholds. These results demonstrate that applying second-order-coherence and photon-counting measurements to Brillouin scattering provides new methods to advance our understanding of Brillouin scattering itself and progress toward quantum-state preparation and characterization of acoustic modes.
\end{abstract}
\date{18/07/2023}% It is always \today, today,

\maketitle

%\tableofcontents

\section{\label{sec:Introduction}Introduction} 

Brillouin--Mandelstam scattering (BMS)---the coupling of two optical fields with an acoustic wave---now celebrates one hundred years since its theoretical elucidation by Brillouin~\cite{Brillouin_1922} and Mandelstam~\cite{Mandelstam_1926} in the early part of the 20th century. The light--matter scattering mechanism is still being actively studied theoretically~\cite{Laude_2015,Sipe_2016,Rakich_2018,Zhang_2023} and experimentally~\cite{Otterstrom_2018,Enzian_2019,Bashan_2022,Xu_2023}, and finds a broad spectrum of applications~\cite{Eggleton_2019, Safavi_2019, Wiederhecker_2019, Wolff_2021} ranging from bulk material characterization~\cite{Sandercock_1972}, determination of acoustic phonon dispersion~\cite{Olsson_2018}, distributed environmental sensing~\cite{Horiguchi_1995}, biophysics and microscopy~\cite{Scarcelli_2015}, as well as the production of high-quality laser light sources~\cite{Gundavarapu_2019}. Indeed, the stimulated amplification of the scattered optical wave, first demonstrated in Ref.~\cite{Chiao_1964}, now provides sub-Hz linewidth and low threshold lasers~\cite{Grudinin_2009,Gundavarapu_2019}. Such narrow-band sources facilitate applications in high-precision interferometry~\cite{Loh_2019}, sensing~\cite{Murray_2022},  Brillouin gyroscopes~\cite{Li_2017}, and information transfer~\cite{Li_2012}. Other recent example advances utilizing Brillouin scattering include microwave synthesis and filtration~\cite{Li_2013, Choudhary_2016}, Brillouin-induced transparency~\cite{Dong_2015, Kim_2015}, and light storage and delay~\cite{Merklein_2018}.

The related field of quantum optics has a similarly long history and now provides a powerful set of theoretical and experimental tools with which to probe and manipulate bosonic systems. In particular, second-order-coherence measurements, introduced by Hanbury Brown and Twiss in 1954~\cite{HanburyBrown_1954} and generalized to quantum fields by Glauber and others~\cite{Glauber_1963}, have since been experimentally employed to great success to characterize the temporal statistics of electromagnetic fields. Second-order-coherence measurements are now utilized to characterize a wide range of optical fields with a prominent example being photon anti-bunching~\cite{Kimble_1977, Grangier_1986, Michler_2000}. Three cases commonly discussed in the quantum optics literature for single-mode second-order coherence at zero-delay $g^{(2)}(0)$  are: $g^{(2)}(0)=1$ for Poissonian coherent light, $g^{(2)}(0)=2$ for thermal light, and $g^{(2)}(0)=0$ for single-photon states.

Building on the successes of quantum optics, quantum optomechanics studies the interaction between optical fields and motional degrees of freedom with an aim to extend quantum optical control to phonons~\cite{Aspelmeyer_2014}. Owing to its engineerability and long phonon coherence times, this growing field of research offers significant potential to advance studies of fundamental physics \cite{Bose_1999,Marshall_2003,Carney_2021}, and develop new technologies such as weak-force sensors~\cite{Kim_2016} and coherent microwave-to-optical transducers~\cite{Higginbotham_2018}. Of particular relevance to this work, photon counting measurements have been utilized within optomechanics to generate nonclassical states of high-frequency vibrations in diamond crystals~\cite{Lee_2012,Fisher_2017,Velez_2019} and photonic-crystal structures~\cite{Riedinger_2016}, perform phonon-counting measurements~\cite{Cohen_2015,Galinsky_2020}, generate mechanical interference fringes \cite{Ringbauer_2018}, observe higher-order phonon correlations~\cite{Patil_2022}, and perform single-phonon addition or subtraction~\cite{Vanner_2013} to thermal states resulting in a doubling of the mean occupation \cite{Enzian_2021_1} and non-Gaussian state generation \cite{Enzian_2021_2,Patel_2021}. Other directions for further experimental pursuit within optomechanics that utilize single-photon counting include quantum state orthogonalization~\cite{Vanner_2013}, kitten-state generation by phonon addition or subtraction to mechanical squeezed states~\cite{Milburn_2016}, growing mechanical superposition states~\cite{Clarke_2018}, and two-mode mechanical entanglement \cite{Borkje_2011,Flayac_2014,KanariNaish_2022}. The properties of backward Brillouin scattering are particularly favourable to implement these types of photon-counting-based protocols and, when combined with the utility of optical second-order-coherence measurements, offer a promising new avenue to generate and explore quantum states with phonons. Notably, quantum optical second-order-coherence measurements are yet to be performed for Brillouin-scattering leaving this avenue unexplored.

This work applies such quantum optical techniques to Brillouin scattering by performing second-order-coherence measurements of Stokes-scattered light across the Brillouin lasing threshold. Utilizing this measurement, we characterize the continuous transition from thermal to coherent statistics and reveal an unexpected feature of super-thermal statistics below the lasing threshold. Our observations are in good agreement with numerical solutions of the nonlinear three-wave-mixing Langevin equations including pump noise, and are well captured by analytic expressions for the instability and lasing threshold derived from the linearized equations of motion. Beyond the characterization of the Brillouin-lasing transition and the observation of this super-thermal feature, photon-counting and second-order-coherence measurements enable several directions for further research and development ranging from device characterization to quantum state preparation of acoustic modes.

\section{\label{sec:TheoryandNumericalSims}Theory and Numerical Simulation}
\begin{figure*}[t]
    \centering
    \includegraphics{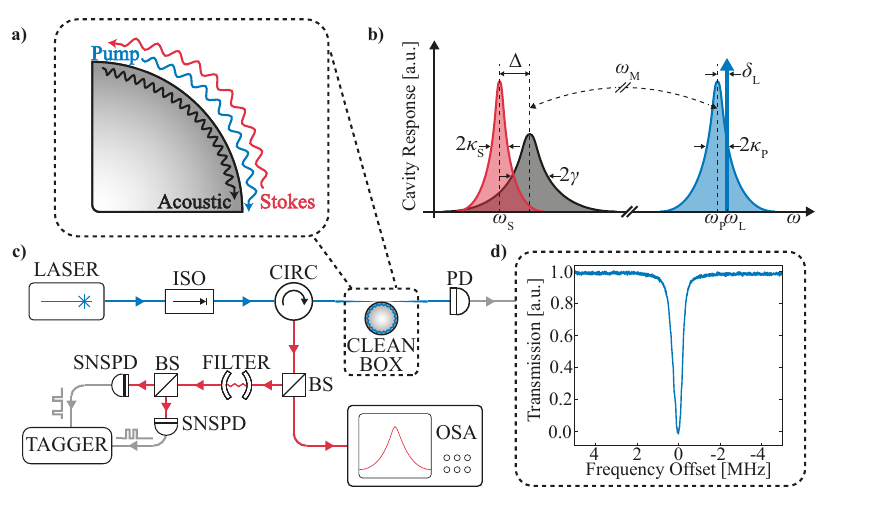}
    \caption{Experimental scheme and optical schematic. \textbf{a)} Three-waving mixing inside a sector of a WGM microresonator. Pump light scatters from a thermally-excited acoustic wave, generating counter-propagating Stokes light. The two optical fields then interfere, creating an intensity beatnote that travels with the speed of sound. \textbf{b)} Resonance condition for the Stokes Brillouin scattering process. A pump mode with angular frequency $\omega_{\text{P}}$ and amplitude decay rate $\kappa_{\text{P}}$ is pumped with a monochromatic laser of angular frequency $\omega_{\text{L}}$ detuned from the pump mode by $\delta_{\text{L}}$, scattering into a lower frequency Stokes mode of the cavity with angular frequency $\omega_{\text{S}}$ and amplitude decay rate $\kappa_{\text{S}}$ off an acoustic wave with angular frequency $\omega_{\text{M}}$ and amplitude decay rate $\gamma$. The detunings $\Delta=\omega_{\text{S}}+\omega_{\text{M}}-\omega_{\text{P}}$ and $\delta_{\text{L}}$ should be small relative to the linewidths to maximise scattering. \textbf{c)} Schematic of the experimental setup. A tuneable pump laser ($1550\,$nm) passes through an isolator (ISO) and circulator (CIRC) and drives an optical microresonator. The transmission spectrum is utilized to locate cavity resonances. The backscattered Stokes signal is separated from the pump with the optical circulator and is divided on a beam splitter (BS). One arm is sent to an optical spectrum analyzer (OSA) to measure the frequency components of the backscattered light and the other is incident on two superconducting-nanowire single-photon detectors (SNSPD). A time-tagger (TAGGER) processes the single-photon-detection signals to determine the second-order coherence. \textbf{d)} Transmission spectrum of the pump mode cavity resonance, normalized to the off-resonant intensity level.}
    \label{fig:Figure1}
\end{figure*}

The nonlinearity of Brillouin-Mandelstam scattering arises from the interplay of the material response to the optical intensity (electrostriction) and the associated modification of the optical field via the acoustic-field-dependent refractive index (photoelasticity). Energy and momentum conservation dictate that two types of optical frequency shift can occur resulting in the enhancement or damping of the acoustic wave. Stokes-scattered light is frequency downshifted from the pump and the acoustic wave is correspondingly amplified, whereas anti-Stokes-scattered light is frequency upshifted and the acoustic wave is damped. In optical cavities, the optical and acoustic fields can be resonantly enhanced or suppressed depending on the phase-matching conditions given by the cavity geometry. With such cavities, pumping one optical mode of a pair of optical resonances separated by approximately the Brillouin shift allows one to resonantly select either the Stokes or anti-Stokes process. This configuration is illustrated in Figure \ref{fig:Figure1}a)-b) for a whispering-gallery-mode (WGM) microresonator. In such a cavity, the BMS interaction can be described by the three-wave-mixing Hamiltonian $H/\hbar = g_0(a_{\text{P}}a_{\text{S}}^{\dagger}b^{\dagger} + a_{\text{P}}^{\dagger}a_{\text{S}}b)$ for an optical pump field $a_{\text{P}}$ of higher frequency, an optical Stokes-scattered field $a_{\text{S}}$ of lower frequency, and an acoustic field $b$, coupled at rate $g_0$. 

The interaction and open-system dynamics of the three fields are well described by the three-wave-mixing Langevin equations. In a rotating frame, the coupled equations for the optical pump, Stokes, and acoustic fields are
\begin{equation}
\label{eqn:langevineqns}
\begin{split}
\Dot{a}_{\text{P}} &= -ig_0a_{\text{S}}b - \kappa_{\text{P}} a_{\text{P}} + \sqrt{2\kappa^{\text{(e)}}_{\text{P}}}a^{\text{(e)}}_{\text{in,P}} +
 \sqrt{2\kappa^{\text{(i)}}_{\text{P}}}a^{\text{(i)}}_{\text{in,P}}\ , \\ 
 \Dot{a}_{\text{S}} &= -ig_0a_{\text{P}}b^{\dagger} - \kappa_{\text{S}} a_{\text{S}} + i\Delta a_{\text{S}} + \sqrt{2\kappa^{\text{(e)}}_{\text{S}}}a^{\text{(e)}}_{\text{in,S}} +
 \sqrt{2\kappa^{\text{(i)}}_{\text{S}}}a^{\text{(i)}}_{\text{in,S}}\ , \\
 \Dot{b} &= -ig_0a_{\text{P}}a_{\text{S}}^{\dagger} - \gamma b + \sqrt{2\gamma}b_{\text{in}} \ ,
\end{split}
\end{equation}
where $\omega_{\text{M}}$, $\omega_{\text{P}}$, and $\omega_{\text{S}}$ are the acoustic, pump mode, and optical Stokes mode angular frequencies, $\Delta=\omega_{\text{S}}+\omega_{\text{M}}-\omega_{\text{P}}$ is the BMS detuning, $\kappa_{\text{P}}$, $\kappa_{\text{S}}$, and $\gamma$ are the total optical pump, Stokes, and acoustic amplitude decay rates, and $\kappa^{\text{(i)}}_{\text{P}}$, $\kappa^{\text{(i)}}_{\text{S}}$ and $\kappa^{\text{(e)}}_{\text{P}}$, 
$\kappa^{\text{(e)}}_{\text{S}}$ are the pump and Stokes intrinsic decay and external coupling rates to the optical environment and output modes, respectively. The complex amplitudes $a^{\text{(i)}}_{\text{in,P}}$, $a^{\text{(e)}}_{\text{in,P}}$, $a^{\text{(i)}}_{\text{in,S}}$, $a^{\text{(e)}}_{\text{in,S}}$ and $b_{\text{in}}$ describe the pump, Stokes and mechanical intrinsic and external inputs including noise. These three coupled nonlinear Langevin equations can be numerically integrated to obtain the full dynamics of the system (see Supplement 1).

To find approximate solutions to Eqns (\ref{eqn:langevineqns}), we can assume a strong, undepletable pump field and linearize the Hamiltonian as the two-mode squeezing interaction $H/\hbar \rightarrow G(a^{\dagger}b^{\dagger} + a b)$ by making the approximation $a_{\text{p}}\rightarrow\alpha$ where $\alpha$ is the intracavity amplitude, which we have taken to be real without loss of generality. Here, $a$ and $b$ are the optical Stokes and acoustic annihilation operators, respectively, and $G = g_0 \alpha$ is the pump-enhanced coupling rate. These equations can be readily solved for weak-coupling ($G \ll \kappa_{\text{S}},\gamma$)  using Fourier transforms to obtain the Brillouin lasing threshold (see Supplement 1), where the round-trip gain in the Stokes mode exceeds the loss. For zero BMS detuning ($\Delta=0$) and critical coupling ($\kappa^{\text{(e)}}_{\text{P}}=\kappa^{\text{(i)}}_{\text{P}}$), the input pump power at the lasing threshold is given by
\begin{equation}
    P_{\text{thresh}} = \frac{\kappa_{\text{S}}\kappa_{\text{P}}\gamma}{g_0^2}\hbar\omega_{\text{L}} \ .
    \label{eq:Pthresh}
\end{equation}
Note that improving the linewidths of any of the three resonances reduces the lasing threshold power.

It can also be shown via the Routh--Hurwitz stability criterion that there exists a lower pump power at which the system becomes unstable to perturbations, arising from the interplay of the pump-enhanced coupling rate  and the two decay rates. This instability occurs for an input pump power of
\begin{equation}
    P_{\text{inst}} = \frac{\kappa_{\text{S}}\kappa_{\text{P}}(\gamma-\kappa_{\text{S}})}{g_0^2}\hbar\omega_{\text{L}} \ ,
    \label{eq:Pinst}
\end{equation}
where we have assumed that $\gamma>\kappa_{\text{S}}$ as is the case in the experimental section that follows. (See also Supplement 1 for a more general discussion and the case where $\gamma<\kappa_{\text{S}}$ which displays an equivalent instability.) Between $P_{\text{inst}}$ and $P_{\text{thresh}}$, we expect to observe increased fluctuations in the optical and mechanical mode amplitudes, manifesting as a measureable change in the second-order coherence of the Stokes field. Using \eqref{eq:Pthresh} and \eqref{eq:Pinst}, the ratio between the onset of instability and the lasing threshold is
\begin{equation}
    \frac{P_{\text{inst}}}{P_{\text{thresh}}} = \frac{\gamma-\kappa_{\text{S}}}{\gamma}\ . 
    \label{eq:Pinst/Pthr}
\end{equation}
Thus, for $\gamma \gg \kappa_{\text{S}}$, the instability and lasing thresholds coincide and, conversely, when $\gamma \simeq \kappa_{\text{S}}$ note that a larger region of instability will be present in the system.

Using Eqs (\ref{eq:Pthresh}-\ref{eq:Pinst/Pthr}), combined with measurements of readily accessible system parameters, allows other key parameters to be estimated. For instance, with knowledge of the pump laser frequency and through measurements of the lasing threshold and the optical and acoustic decay rates, $P_{\text{inst}}$ and $g_0$ can be determined. Conversely, $P_{\text{thresh}}$ and $P_{\text{inst}}$ in conjunction with the optical linewidths can be used to determine $\gamma$ and $g_0$. Such estimates are valuable for system characterization and provide an additional method to dyne-detection-based approaches.

Of key interest to this work, the second-order coherence of the Stokes-scattered light will have a continuous transition between the two regimes well below and above the Brillouin lasing threshold. The second-order coherence at zero delay~\cite{Glauber_1963} is defined as
$
     g^{(2)}(0) = \left\langle n(n-1)\right\rangle/\left\langle n\right\rangle^2,
$
where $n$ is the photon number operator of the optical Stokes mode, and may be understood for this physical system by considering the differing situations under which the Stokes field is produced. Below threshold, pump photons scatter from thermally excited phonons and leave the cavity, either through external coupling or losses, before a coherent amplitude can build up and therefore display thermal correlations and a second-order coherence of $g^{(2)}(0)=2$. Above threshold, however, scattering occurs at a large enough rate that cavity losses are overcome by Brillouin gain, constituting a positive feedback loop as the mechanical mode is also amplified by the scattering process. In this case, pump photons are coherently scattered into the Stokes mode and the output light displays Poissonian lasing statistics with $g^{(2)}(0)=1$. 

In the intermediate region between $P_{\text{inst}}$ and $P_{\text{thresh}}$ however, the nonlinear nature of the Brillouin interaction becomes apparent: both super-thermal and sub-thermal statistics are observed as a result of the instability. Super-thermal statistics arise from fluctuations in the pump and acoustic fields causing the system to begin to lase only for the pump field to be depleted, causing a ``flickering'' in Stokes-scattered power, which manifests as an increase in the second-order coherence on top of thermal light. This pump depletion occurs as the self-reinforcing amplification of the coupled Stokes and acoustic modes exceeds the rate at which energy is coupled into the pump mode from the drive laser. Sub-thermal statistics arise from the system slipping increasingly over the lasing threshold and accumulating a coherent optical amplitude. The interplay of these effects first raise $g^{(2)}(0)$ above 2, followed by a monotonic decrease to Poissonian statistics as the pump power is increased above the lasing threshold. To fully capture the non-linear three-wave mixing, Eqns~(\ref{eqn:langevineqns}) are numerically integrated via the Euler method, neglecting optical vacuum noise terms and assuming room-temperature thermal acoustic fluctuations (see Supplement 1). Importantly, this model includes classical noise on the intracavity pump amplitude. The resultant Brillouin coupling fluctuations are introduced via a complex noise term in the pump field which models fluctuations in the intracavity amplitude. The results of this numerical simulation are discussed together with the experimental results below.

% sources that affect the intracavity pump photon number are included. \textcolor{red}{Such noise sources include pump laser power and frequency fluctuations, and variations in the cavity coupling conditions.}

%In the intermediate region between these two regimes, an analytic expression for $g^{(2)}(0)$ is not known.

% $P_{\text{inst}}$ and $P_{\text{thresh}}$, 
% both super-thermal and subthermal statistics are observed as a result of the instability. The former arise from the fluctuations in Stokes intensity which manifests as additional second-order coherence on top of thermal light and the latter from the system slipping increasingly over the lasing threshold and containing a larger coherent component. The interplay of these effects first raise $g^{(2)}(0)$ above 2, followed by a monotonic decrease to Poissonian statistics as the pump power is increased above the lasing threshold.

\section{\label{sec:ExperimentalScheme}Experimental Setup}

A schematic of the experimental setup is shown in Figure \ref{fig:Figure1}c). Pump light from a $1550\,$nm widely-tunable laser is sent through an isolator and optical circulator and then evanescently coupled into a 280 \textmu m diameter silica  microsphere resonator via a tapered fiber. The tapered fiber and silica microsphere were made using a hydrogen-flame tapering rig and a fusion-arc splicer, respectively. The transmitted signal is sent onto a photodiode to characterise the optical resonances of the microsphere, one of which is plotted in Fig. \ref{fig:Figure1}d). A pair of optical resonances (intensity decay rates $2\kappa_{\text{P}}/2\pi=1.1\,$MHz and $2\kappa_\text{{S}}/2\pi=3.2\,$MHz) separated by the backward Brillouin acoustic frequency, $\omega_{\text{M}}/2\pi\simeq11\,$GHz, are used to selectively drive the Stokes interaction. The pump laser is passively locked to the higher frequency resonance of the pair utilizing optical nonlinearities~\cite{Carmon_2004} and was maintained at a fixed detuning from the pump resonance by $\delta_{\text{L}}/2\pi=200$ kHz.

The experiment was performed with input pump powers at the tapered fiber ranging from approximately 40 to 100 \textmu W resulting in intracavity powers of approximately 4 to 10 W. Measurements were made at room temperature (295 K), corresponding to a mean thermal acoustic occupation of $\Bar{n}_{\text{th}}=558$. At this temperature, we observe the acoustic intensity decay rate (Brillouin linewidth) to be $2\gamma/2\pi=27$~MHz, which is consistent with previous measurements and understanding of amorphous silica~\cite{Boyd1992,LeFloch_2003_1}.

The backscattered Stokes signal is coupled out of the resonator via the same tapered optical fiber and is separated from the pump field using an optical circulator. The signal field is identified and measured using an optical spectrum analyzer (OSA) and the second-order-coherence measurements are performed with photon counters. To implement the $g^{(2)}(0)$ measurement, two fiber-based Fabry-Perot filters with intensity full-widths-at-half-maxima of $120\,$MHz and a free-spectral range of $25\,$GHz are used to remove spurious pump light, then the filtered light is sent through another 50:50 beam splitter prior to detection by a pair of superconducting-nanowire single-photon detectors (SNSPDs). The signals from these detectors are processed by a time tagger from which the second-order coherence ($g^{(2)}(0)$) is obtained. The SNSPDs are operated with a reduced bias current to avoid latching in the lasing regime which yielded a dark count rate of approximately $1\,$s\textsuperscript{-1}.

\section{\label{sec:Results}Results and Discussion}

\begin{figure*}[t]
    \centering
    \includegraphics[width=0.6\textwidth]{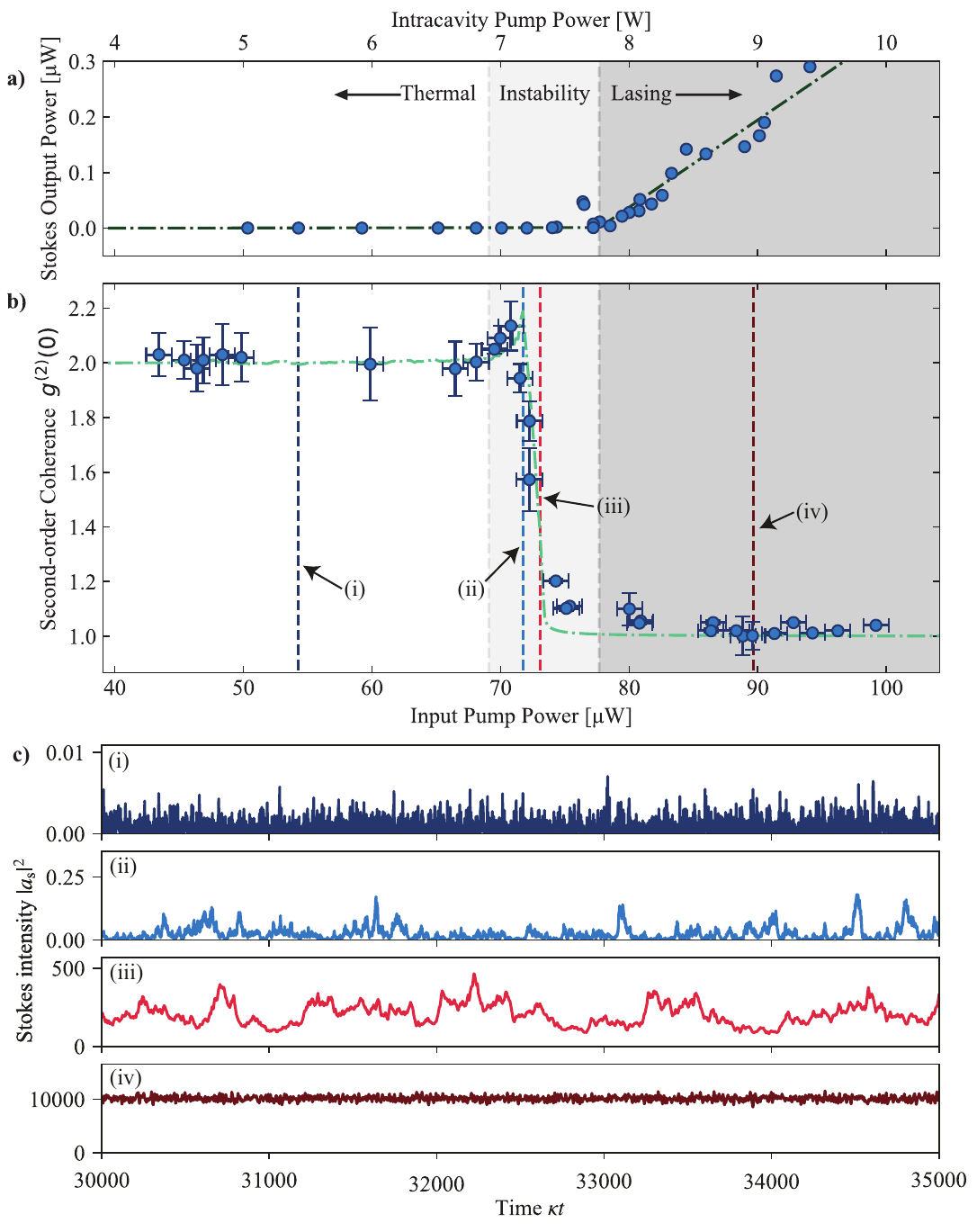}
    \caption{Plots of \textbf{a)} Stokes-scattered output power, \textbf{b)} second-order coherence $g^{(2)}(0)$ against input pump power across the Brillouin lasing transition, and \textbf{c)} numerically simulated Stokes intensity time series from which numerical values of $g^{(2)}(0)$ are calculated. In \textbf{a)} and \textbf{b)}, filled circles are experimental points and dark and light green dotted lines are linear fits of the Stokes-scattered power and numerical simulations, respectively. The build-up factor between the input power (lower horizontal axis) and the intracavity power (upper horizontal axis) is $10.1\times10^{4}$. Both \textbf{a)} and \textbf{\textbf{b)}} are divided into three regions corresponding to thermal light (white background), instability (light grey background), and above the lasing threshold (dark grey background). In \textbf{a)}, the change in slope of the Stokes output power indicates the Brillouin lasing threshold and the error bars in these measurements are similar in size to the circles. In \textbf{b)}, the second-order coherence of the Stokes field behaves distinctly for each of these three regions: for low powers, the field displays thermal statistics, then as the power is increased towards the lasing transition, super-thermal statistics are observed followed by a monotonic decrease to Poissonian light above the lasing threshold. The lasing threshold is determined from the measured kink in the Stokes-scattered power and the instability region is determined from ~\eqref{eq:Pinst/Pthr} using experimentally determined parameters in good agreement with the observed onset of super-thermal statistics. The dashed lines (i-iv) in \textbf{b)} correspond to the labelled time series in \textbf{c)}: (i) far below threshold (dark blue), the Stokes field displays thermal intensity fluctuations with very little coherent amplitude, (ii) within the region of instability (light blue), large spikes in intensity are visible as the Stokes mode begins to “flicker”, displaying super-thermal statistics, (iii) below the lasing threshold (light red), the flickering persists but is now on top of a coherent intensity, resulting in a second-order-coherence between thermal and Poissonian statistics, and (iv) far above threshold (dark red), intensity fluctuations are small relative to the strong coherent intensity and the light displays Poissonian statistics.}
    \label{fig:G2Plot}
\end{figure*}

The backscattered Stokes output power as measured by the OSA is plotted against the input pump power in Fig.~\ref{fig:G2Plot}a). For low input pump powers, the Stokes-scattered signal is weak and composed of thermally scattered light. As the power is increased, the lasing threshold is identified at $78.0$~\textmu W input power where a clear discontinuity in the gradient is observed as the Stokes optical mode begins to lase. Linear fits to the thermal and lasing regions are plotted and give good agreement with the power measurements.

% and the output power increases linearly with pump strength with a lasing efficiency of 1.54\%.

The measured second-order coherence $g^{(2)}(0)$ is plotted in Fig.~\ref{fig:G2Plot}b) over the same range of input pump powers, together with numerical solutions of $g^{(2)}(0)$ calculated from the nonlinear three-wave-mixing Langevin equations. For low input powers, the second-order coherence displays thermal correlations, i.e. $g^{(2)}(0)=2$, as pump light scatters off thermally excited acoustic waves, which imprint their thermal correlations onto the optical Stokes field. As the input power approaches the lasing threshold, a continuous transition in second-order coherence is observed, which notably displays an increase in $g^{(2)}(0)$ above 2, corresponding to super-thermal statistics. This behaviour arises from the interplay between intensity fluctuations, and the relative fraction of time the system slips in and out of lasing. Input powers just above the point of instability result in large intensity fluctuations as the Stokes mode begins lasing, temporarily depleting the pump field and resulting in an immediate drop in Stokes intensity; as a result, the observed intensity correlations are strongly bunched, or super-thermal. As the system is pumped more strongly, it spends a greater proportion of time lasing and thus contains a greater coherent component with $g^{(2)}(0)$ approaching 1 as the stimulated BMS lasing threshold is exceeded. In Fig.~\ref{fig:G2Plot}c), time series of the numerically computed Stokes intensity are plotted for the powers labelled (i-iv).

For the lasing threshold input power observed $P_{\text{thresh}}=78$~\textmu W, corresponding to approximately $10^8$ intracavity photons and a pump-enhanced coupling rate of $G/2\pi=3.86$~MHz, the onset of the instability and its width are in good agreement with Eq.~(\ref{eq:Pinst}), which gives $P_{\text{inst}}=69\,$\textmu W. Additionally, following the procedure we describe to estimate system parameters using Eqns (\ref{eq:Pthresh}-\ref{eq:Pinst/Pthr}), we obtain $g_0/2\pi\simeq309$~Hz, which is consistent with simulation and other experimental work in similar silica microresonators~\cite{Enzian_2019}. From the experimental parameters used in the numerical simulations, summarised in Supplement 1, we obtain good agreement between the observed and simulated second-order coherence (c.f. Fig.~\ref{fig:G2Plot}b)) for noise on the intracavity intensity corresponding to a 1 part in 50 variation. We expect this variation to originate from pump frequency fluctuations; for a pump laser detuned by 200~kHz from the centre of the 1.1~MHz linewidth resonance, this intensity variation arises from a pump linewidth of approximately $15$~kHz, consistent with the frequency noise of the laser used.

\section{Conclusion and Outlook}
\label{sec:DiscussionConclusion}

Using a silica microsphere resonator and single-photon detectors, we have characterized the second-order coherence of the optical Stokes field across the Brillouin lasing transition. We observe that the zero-delay second-order coherence $g^{(2)}(0)$ continuously transitions from a value of 2 well below the lasing threshold to a value of 1 above the threshold. This observation is in line with the conceptual understanding that the Stokes field changes from being thermal to coherent as it is driven across the threshold. During this transition, however, for a range of input powers below the threshold, there is a region of instability where the second-order coherence changes between these two regimes and can take on a wider range of values. Indeed, within this region of instability we observe super-thermal statistics, i.e. $g^{(2)}(0) > 2$, which arises due to a ``flickering'' of the intensity as the Stokes mode slips in and out of lasing. This behaviour of the second-order coherence across the lasing threshold is not describable by a linearized model but is well captured by numerical solutions of the full three-wave-mixing process that gives the transient pump depletion in the unstable region. The input-power range of this unstable region increases when the acoustic and optical decay rates are similar and the magnitude of the super-thermal statistics depends on the fluctuations of the optical pump field. These findings highlight the need to carefully consider this instability for Brillouin Stokes scattering applications including pulsed and continuous Brillouin light-matter two-mode squeezing protocols, and low-power Brillouin-lasing applications operating just above threshold. To the best of our knowledge, this is the first application of second-order-coherence measurements to Brillouin--Mandelstam scattering and the first characterization of the second-order coherence across the Brillouin lasing transition. More broadly, this experimental work advances the use of photon counting for Brillouin--Mandelstam scattering, enabling new approaches to system characterization and helps pave a foundation for research and development that pursues quantum science and technology with acoustic modes.

\section*{Funding Information}
This project was supported by UK Research and Innovation (MR/S032924/1), the Engineering and Physical Sciences Research Council (EP/T031271/1, EP/P510257/1), the Science and Technology Facilities Council (ST/W006553/1), the Royal Society, the Aker Scholarship, and the EU Horizon 2020 Program (847523 “INTERACTIONS”).

\section*{Acknowledgments}
The authors would like to acknowledge useful discussions with Mickael Chan, Lydia Kanari-Naish, W. Steven Kolthammer, Gerard J. Milburn, and Michael Woodley.

%\nocite{*}

\bibliography{refs}% Produces the bibliography via BibTeX.

%%%%%%%%%%%%%%%%%%%%%%%%%%%%%%%%%%%%%%%%%%%%%%%%%%%%%%%%%%%%%%%%%
%%%%% SUPPLEMENTARY %%%%%
%%%%%%%%%%%%%%%%%%%%%%%%%%%%%%%%%%%%%%%%%%%%%%%%%%%%%%%%%%%%%%%%%

\onecolumngrid
\clearpage
\setcounter{equation}{0}
\setcounter{figure}{0}
\setcounter{section}{0}
\def\theequation{S\arabic{equation}}
\def\thefigure{S\arabic{figure}}
\def\thetable{S\arabic{table}}
\def\thesection{S\arabic{section}}

\begin{center}
\textbf{\Large{Supplementary Material}}
\end{center}
\section{Heisenberg-Langevin Equations of Motion}

The Heisenberg-Langevin equations of motion describe the time evolution of the optical and acoustic field operators, accounting for unitary dynamics, dissipation, and thermal and quantum fluctuations arising from coupling to their environments. For the Brillouin-optomechanical system of interest here, the equations of motion for the optical pump ($a_{\text{P}}$/$a^{\dagger}_{\text{P}}$), Stokes ($a_{\text{S}}$/$a^{\dagger}_{\text{S}}$), and mechanical ($b$/$b^{\dagger}$) modes read
\begin{align}
\label{eqn:HLeqns}
\begin{split}
\Dot{a}_{\text{P}} &= -i\left[a_{\text{P}},H/\hbar\right] - \kappa_{\text{P}} a_{\text{P}} + \sqrt{2\kappa^{\text{(e)}}_{\text{P}}}a^{\text{(e)}}_{\text{in,P}} +
 \sqrt{2\kappa^{\text{(i)}}_{\text{P}}}a^{\text{(i)}}_{\text{in,P}}\ , \\ 
 \Dot{a}_{\text{S}} &= -i\left[a_{\text{S}},H/\hbar\right] - \kappa_{\text{S}} a_{\text{S}} + \sqrt{2\kappa^{\text{(e)}}_{\text{S}}}a^{\text{(e)}}_{\text{in,S}} +
 \sqrt{2\kappa^{\text{(i)}}_{\text{S}}}a^{\text{(i)}}_{\text{in,S}}\ , \\
 \Dot{b} &= -i\left[b,H/\hbar\right] - \gamma b + \sqrt{2\gamma}b_{\text{in}} \ ,
\end{split}
\end{align}
where $\kappa_{\text{P}}$, $\kappa_{\text{S}}$, and $\gamma$ are the total optical pump, Stokes, and mechanical amplitude decay rates and $\kappa^{\text{(i)}}_{\text{P}}$, $\kappa^{\text{(i)}}_{\text{S}}$, and $\kappa^{\text{(e)}}_{\text{P}}$, 
$\kappa^{\text{(e)}}_{\text{S}}$ are the pump and Stokes intrinsic decay, and external coupling rates to the optical environment and output modes, respectively. The operators $a^{\text{(i)}}_{\text{in,P}}$, $a^{\text{(e)}}_{\text{in,P}}$, $a^{\text{(i)}}_{\text{in,S}}$, $a^{\text{(e)}}_{\text{in,S}}$ and $b_{\text{in}}$ describe the pump, Stokes and mechanical intrinsic and external inputs including noise and it has been assumed that the mechanical field is not subjected to an external drive. The nonlinear three-waving mixing Hamiltonian $H$ describing the Stokes Brillouin-Mandelstam interaction is given by 
\begin{equation}
    H/\hbar = \omega_{\text{P}} a^{\dagger}_{\text{P}}a_{\text{P}} + \omega_{\text{S}} a^{\dagger}_{\text{S}}a_{\text{S}} + \omega_{\text{M}}b^{\dagger}b + g_0\left(a^{\dagger}_{\text{P}}a_{\text{S}}b + a_{\text{P}}a^{\dagger}_{\text{S}}b^{\dagger}\right) \ .
    \label{eq:3waveHamiltonian}
\end{equation}
Here, $\omega_{\text{P/S/M}}$ are the pump, Stokes and mechanical angular frequencies, respectively, and $g_0$ is the three-wave-mixing coupling rate.

Moving into a rotating frame defined by $a_{\text{P}}\rightarrow a_{\text{P}}e^{i\omega_{\text{P}}t}$, $a_{\text{S}}\rightarrow a_{\text{S}}e^{i(\omega_{\text{P}}-\omega_{\text{M}})t}$, and $b\rightarrow be^{i\omega_{\text{M}}t}$ (and similarly for the input noise operators), the Hamiltonian becomes
\begin{equation}
    H/\hbar = -\Delta a^{\dagger}_{\text{S}}a_{\text{S}} + g_0\left(a^{\dagger}_{\text{P}}a_{\text{S}}b + a_{\text{P}}a^{\dagger}_{\text{S}}b^{\dagger}\right) \ ,
    \label{eq:3waveHamiltoniandetuning}
\end{equation}
where $\Delta=\omega_{\text{P}}-\omega_{\text{S}}-\omega_{\text{M}}$ is the optomechanical detuning. After evaluating the commutators in Eq.~(\ref{eqn:HLeqns}), we obtain the full non-linear equations of motion for the full system
\begin{align}
\label{eqn:HLeqnsfull}
\begin{split}
\Dot{a}_{\text{P}} &= -ig_0a_{\text{S}}b - \kappa_{\text{P}} a_{\text{P}} + \sqrt{2\kappa^{\text{(e)}}_{\text{P}}}a^{\text{(e)}}_{\text{in,P}} +
 \sqrt{2\kappa^{\text{(i)}}_{\text{P}}}a^{\text{(i)}}_{\text{in,P}}\ , \\ 
 \Dot{a}_{\text{S}} &= i\Delta a_{\text{S}}-ig_0a_{\text{P}}b^{\dagger} - \kappa_{\text{S}} a_{\text{S}} + \sqrt{2\kappa^{\text{(e)}}_{\text{S}}}a^{\text{(e)}}_{\text{in,S}} +
 \sqrt{2\kappa^{\text{(i)}}_{\text{S}}}a^{\text{(i)}}_{\text{in,S}}\ , \\
 \Dot{b} &= -ig_0a_{\text{P}}a_{\text{S}}^{\dagger} - \gamma b + \sqrt{2\gamma}b_{\text{in}} \ .
\end{split}
\end{align}
These differential equations can then be solved numerically as is discussed in Section \ref{sec:numerics} below.

\subsection{Linearized Approximation}

Although the second-order coherence measurements are not describable by a linearized model and require the full three-wave-mixing interaction including pump depletion, we can gain insight into the lasing threshold and instability using a linearized approximation. We consider a strong (coherent) pump and use the mapping $a_{\text{P}}\rightarrow\alpha_{\text{P}}$ where $\alpha_{\text{P}}$ is a complex amplitude which we assume to be real without any loss of generality. The linearized Hamiltonian is then
\begin{equation}
    H_{\text{lin}}/\hbar = -\Delta a^{\dagger}_{\text{S}}a_{\text{S}} + G\left(a_{\text{S}}b + a^{\dagger}_{\text{S}}b^{\dagger}\right) \ ,
    \label{eq:2waveHamiltonian}
\end{equation}
where $G=g_0\alpha_{\text{P}}$ is the pump-enhanced optomechanical coupling rate and the interaction has now taken the form of a two-mode-squeezing Hamiltonian.
The linearized Heisenberg-Langevin equations for the Stokes and mechanical modes then read
\begin{align}
\label{eqn:HLeqnslinear}
\begin{split}
 \Dot{a}_{\text{S}} &= i\Delta a_{\text{S}} - iG b^{\dagger} - \kappa_{\text{S}} a_{\text{S}} + \sqrt{2\kappa^{\text{(e)}}_{\text{S}}}a^{\text{(e)}}_{\text{in,S}} +
 \sqrt{2\kappa^{\text{(i)}}_{\text{S}}}a^{\text{(i)}}_{\text{in,S}}\ , \\
 \Dot{b} &= - iG a^{\dagger}_{\text{S}} - \gamma b  + \sqrt{2\gamma}b_{\text{in}} \ , 
\end{split}
\end{align}
where the Stokes interaction correlates the optical annihilation operator with the mechanical creation operator of the system and vice versa. We therefore take the adjoint of the mechanical field to write this pair of equations in matrix form
\begin{gather}
 \begin{pmatrix} \Dot{a}_{\text{S}} \\ \Dot{b^{\dagger}} \end{pmatrix}
 =
  \begin{pmatrix}
   - \kappa_{\text{S}} + i\Delta &
   -iG \\
   iG &
   -\gamma  
   \end{pmatrix}
   \begin{pmatrix}
    a_{\text{S}} \\
    b^{\dagger}
   \end{pmatrix}
   +
   \begin{pmatrix}
\sqrt{2\kappa^{\text{(e)}}_{\text{S}}}a^{\text{(e)}}_{\text{in,S}} +
 \sqrt{2\kappa^{\text{(i)}}_{\text{S}}}a^{\text{(i)}}_{\text{in,S}} \\
   \sqrt{2\gamma}b_{\text{in}}^{\dagger}
   \end{pmatrix} \ .
\label{eq:CoupledMatrixEq}
\end{gather}
These equations of motion can now be solved using transform methods for weak coupling and examined for the stability of their solutions.

\subsection{Determination of the Brillouin Lasing Threshold}

In order to derive the Brillouin lasing threshold, we first take the Fourier transform of Eq. (\ref{eq:CoupledMatrixEq}), obtaining
\begin{gather}
 -i\omega\begin{pmatrix} \Tilde{a}_{\text{S}} \\ \Tilde{b}^{\dagger} \end{pmatrix}
 =
  \begin{pmatrix}
   - \kappa_{\text{S}} + i\Delta &
   -iG \\
   iG &
   -\gamma  
   \end{pmatrix}
   \begin{pmatrix}
    \Tilde{a_{\text{S}}} \\
    \Tilde{b}^{\dagger}
   \end{pmatrix}
   +
   \begin{pmatrix}
\sqrt{2\kappa^{\text{(e)}}_{\text{S}}}\Tilde{a}^{\text{(e)}}_{\text{in,S}} +
 \sqrt{2\kappa^{\text{(i)}}_{\text{S}}}\Tilde{a}^{\text{(i)}}_{\text{in,S}}\\
   \sqrt{2\gamma}\Tilde{b}_{\text{in}}^{\dagger}
   \end{pmatrix} \ ,
\label{eq:CoupledMatrixFTEq}
\end{gather}
where $\Tilde{a}_{\text{S}}$, $\Tilde{b}^{\dagger}$, $\Tilde{a}_{\text{in,S}}$ and $\Tilde{b}_{\text{in}}^{\dagger}$ are the Fourier transformed duals of $a_{\text{S}}$, $b^{\dagger}$, $a_{\text{in,S}}$ and $b_{\text{in}}^{\dagger}$. Rearranging and finding the inverse of the coefficient matrix of $\Tilde{a}_{\text{S}}$ and $\Tilde{b}^{\dagger}$, we find
\begin{gather}
 \begin{pmatrix} \Tilde{a}_{\text{S}} \\ \Tilde{b}^{\dagger} \end{pmatrix}
 =
 \frac{1}{(\kappa_{\text{S}} - i(\omega + \Delta))(\gamma - i\omega) - G^2}
  \begin{pmatrix}
    \gamma - i\omega &
   -iG \\
   iG &
   \kappa_{\text{S}}  -i(\omega +\Delta)
   \end{pmatrix}
   \begin{pmatrix}
\sqrt{2\kappa^{\text{(e)}}_{\text{S}}}\Tilde{a}^{\text{(e)}}_{\text{in,S}} +
 \sqrt{2\kappa^{\text{(i)}}_{\text{S}}}\Tilde{a}^{\text{(i)}}_{\text{in,S}}\\
   \sqrt{2\gamma}\Tilde{b}_{\text{in}}^{\dagger}
   \end{pmatrix} \ .
\label{eq:CoupledMatrixInverseEq}
\end{gather}
For $\lvert\omega+\Delta\rvert\ll\kappa_{\text{S}}$, this can be simplified to
\begin{gather*}
 \begin{pmatrix} \Tilde{a}_{\text{S}} \\ \Tilde{b}^{\dagger} \end{pmatrix}
 =
 \frac{1}{\kappa_{\text{S}}(\gamma_{\text{eff}}-i\omega)}
  \begin{pmatrix}
    \gamma - i\omega &
   -iG \\
   iG &
   \kappa_{\text{S}}  -i(\omega +\Delta)
   \end{pmatrix}
   \begin{pmatrix}
\sqrt{2\kappa^{\text{(e)}}_{\text{S}}}\Tilde{a}^{\text{(e)}}_{\text{in,S}} +
 \sqrt{2\kappa^{\text{(i)}}_{\text{S}}}\Tilde{a}^{\text{(i)}}_{\text{in,S}} \\
   \sqrt{2\gamma}\Tilde{b}_{\text{in}}^{\dagger}
   \end{pmatrix} \ ,
\label{eq:CoupledMatrixInverseEq}
\end{gather*}
where $\gamma_{\text{eff}}$ is the effective mechanical linewidth, given by
\begin{equation}
    \gamma_{\text{eff}} = \gamma\left(1-\frac{G^2}{\gamma\kappa_{\text{S}}}\right) \ ,
\end{equation}
which goes to 0 as $G^2$ approaches $\gamma\kappa_{\text{S}}$. This is equivalent to the optomechanical cooperativity ($C=G^2/\kappa\gamma$) exceeding unity, and the input power for which the mechanical linewidth vanishes serves as a definition of the lasing threshold. The pump-enhanced coupling rate $G = g_0 \alpha_{\text{P}} = g_0 \sqrt{n_\text{cav}}$, where $n_\text{cav}$ is the intracavity pump photon number, is related to the input power $P_{\text{in}}$ via
\begin{equation}
    \label{eq:intracavphotonnumber}
    n_{\text{cav}} = \frac{2\kappa^{\text{(e)}}_{\text{P}}}{\kappa_{\text{P}}^2}\left(\frac{\kappa_{\text{P}}^2}{\kappa_{\text{P}}^2+\delta_{\text{L}}^2}\right) \frac{P_{\text{in}}}{\hbar\omega_{\text{P}}}\ ,
\end{equation}
where $\kappa^{\text{(e)}}_{\text{P}}$ is the external amplitude decay rate and $\delta_{\text{L}}$ is the detuning of the pump laser from the pump mode resonance. Substituting this into $G^2=\gamma\kappa_{\text{S}}$ and rearranging for the lasing threshold power $P_{\text{thresh}}$, we arrive at
\begin{equation}
    P_{\text{thresh}} = \frac{\gamma\kappa_{\text{P}}^2\kappa_{\text{S}}}{2\kappa^{\text{(e)}}_{\text{P}}g_0^2}\left(\frac{\kappa_{\text{P}}^2+\delta_{\text{L}}^2}{\kappa_{\text{P}}^2}\right)\hbar\omega_{\text{P}}\ .
\end{equation}

\begin{table}[h]
    \centering
    \caption{Order-of-magnitude example parameters for silica microspheres at the telecommunications wavelength of 1550 nm.}
    \begin{tabular}{|c|c|c|c|c|c|}
         \hline
         \textbf{Parameter} &
         $2\gamma/2\pi$ & $2\kappa_{\text{P}}/2\pi=2\kappa_{\text{S}}/2\pi$ & $g_0/2\pi$ & $\delta_{\text{L}}/2\pi$ & $\hbar\omega_{\text{P}}$\\
         \hline
         \textbf{Value} & 
         $10^7\,$Hz & $10^6\,$Hz & $10^2\,$Hz & $10^5\,$Hz & $10^{-19}\,$J \\
         \hline
    \end{tabular}
    \label{tab:ExampleParameters}
\end{table}
For critical coupling ($\kappa_{\text{P}}=2\kappa^{\text{(e)}}_{\text{P}}$) and for the order-of-magnitude parameters given in Table \ref{tab:ExampleParameters}, we predict a Brillouin lasing threshold of $P_{\text{thresh}}\simeq100\,$\textmu W. 
Then, again assuming critical coupling and for resonant pumping ($\delta_{\text{L}}=0$), we obtain the equation given in the main text
\begin{equation}
    P_{\text{thresh}} = \frac{\gamma\kappa_{\text{P}}\kappa_{\text{S}}}{g_0^2}\hbar\omega_{\text{P}} \ ,
\end{equation}
which represents the lowest input power with which to achieve lasing.

\subsection{Determination of the Brillouin Lasing Instability}

We can also examine the eigenvalue spectrum of the coupling matrix in Eq.~(\ref{eq:CoupledMatrixEq}) to determine parameters for which the eigenvalues have a positive real part corresponding to exponential growth, i.e. examine the Routh-Hurwitz stability criterion. Calculating the eigenvalues of the coupling matrix $\lambda_{\pm}$, we obtain
\begin{equation}
    \lambda_{\pm} = \frac{-\kappa_{\text{S}}-\gamma-i\Delta}{2}\pm
    \frac{\kappa_{\text{S}}-\gamma-\lvert\Delta\rvert}{2}\sqrt{1+\frac{4G^2 + 2\Delta\left(\kappa_{\text{S}}-\gamma\right)(1+i)}{(\kappa_{\text{S}}-\gamma)^2-\Delta^2}} \ .
\end{equation}
Assuming detunings much smaller than amplitude decay rates ($\Delta/\kappa_{\text{S}},\Delta/\gamma\ll1$), we Taylor expand the square root and drop all terms $O(\Delta^2)$ to obtain
\begin{equation*}
    \lambda_{\pm} = \frac{-\kappa_{\text{S}}-\gamma-i\Delta}{2}\pm
    \left(
    \frac{\kappa_{\text{S}}-\gamma-\lvert\Delta\rvert}{2} 
    + \frac{(\kappa_{\text{S}}-\gamma-\lvert\Delta\rvert)G^2)}{(\kappa_{\text{S}}-\gamma)^2}
    + \frac{\Delta(\kappa_{\text{S}}-\gamma)}{2(\kappa_{\text{S}}-\gamma)}
    +\frac{i\Delta(\kappa_{\text{S}}-\gamma)}{2(\kappa_{\text{S}}-\gamma)}
    \right) \ .
\end{equation*}
In this experiment, the mechanical damping rate $\gamma$ exceeds the cavity decay rates $\kappa_{\text{S}}$, $\kappa_{\text{P}}$. We therefore first concern ourselves with $\lambda_{-}$ as it is able to take positive real values for small $\Delta$. Grouping its real and imaginary parts, we see that
\begin{equation}
    \lambda_{-} = \lambda_{\mathcal{R}-} + i\lambda_{\mathcal{I}-} = -\kappa_{\text{S}} - \frac{G^2}{(\kappa_{\text{S}}-\gamma)} + \frac{\lvert\Delta\rvert G^2}{(\kappa_{\text{S}}-\gamma)^2} + i\lambda_{\mathcal{I}-} \ ,
\end{equation}
which, for the condition $\lambda_{\mathcal{R}-}>0$, implies 
\begin{equation*}
    G^2 > \frac{\kappa_{\text{S}}(\gamma-\kappa_{\text{S}})}{1+\lvert\Delta\rvert/(\gamma-\kappa_{\text{S}})} \ .
\end{equation*}
Using Eq. (\ref{eq:intracavphotonnumber}), we can express this instability threshold in terms of input pump power $P_{\text{inst}}$ as
\begin{equation}
    P_{\text{inst}} = \frac{\kappa_{\text{P}}^2(\gamma-\kappa_{\text{S}})\kappa_{\text{S}}}{2\kappa^{\text{(e)}}_{\text{P}}g_0^2[1+\lvert\Delta\rvert/(\gamma-\kappa_{\text{S}})]}
    \left(\frac{\kappa_{\text{P}}^2+\delta_{\text{L}}^2}{\kappa_{\text{P}}^2}\right)\hbar\omega_{\text{P}} \ ,
\end{equation}
which reduces to 
\begin{equation}
    P_{\text{inst}} = \frac{\kappa_{\text{P}}(\gamma-\kappa_{\text{S}})\kappa_{\text{S}}}{g_0^2}\hbar\omega_{\text{P}} \ ,
    \label{eq:Pinst_supp}
\end{equation}
for critical coupling, and zero optomechanical and pump detuning i.e. $\Delta = \delta_{\text{L}} = 0$. For the order-of-magnitude parameters in Table \ref{tab:ExampleParameters} above, the onset of instability is at $P_{\text{inst}}\simeq90\,$\textmu W.

Using the same procedure as above, for systems where the optical decay rates exceed those of the mechanics, the instability threshold is instead given by
\begin{equation}
    P_{\text{inst}} = \frac{\kappa_{\text{P}}^2\gamma(\kappa_{\text{S}}-\gamma)}{2\kappa^{\text{(e)}}_{\text{P}}g_0^2[1-\lvert\Delta\rvert/(\kappa_{\text{S}}-\gamma)]}
    \left(\frac{\kappa_{\text{P}}^2+\delta_{\text{L}}^2}{\kappa_{\text{P}}^2}\right)\hbar\omega_{\text{P}} \ ,
\end{equation}
which is obtained by taking the other branch, $\lambda_{+}$. It is clear that the system is therefore similarly unstable under a change of the damping hierarchy and this effect should therefore also be observable for systems with optical losses exceeding mechanical decay rates.

\subsection{Numerical Simulation}
\label{sec:numerics}

\begin{figure}
    \centering
    \includegraphics[width=0.8\textwidth]{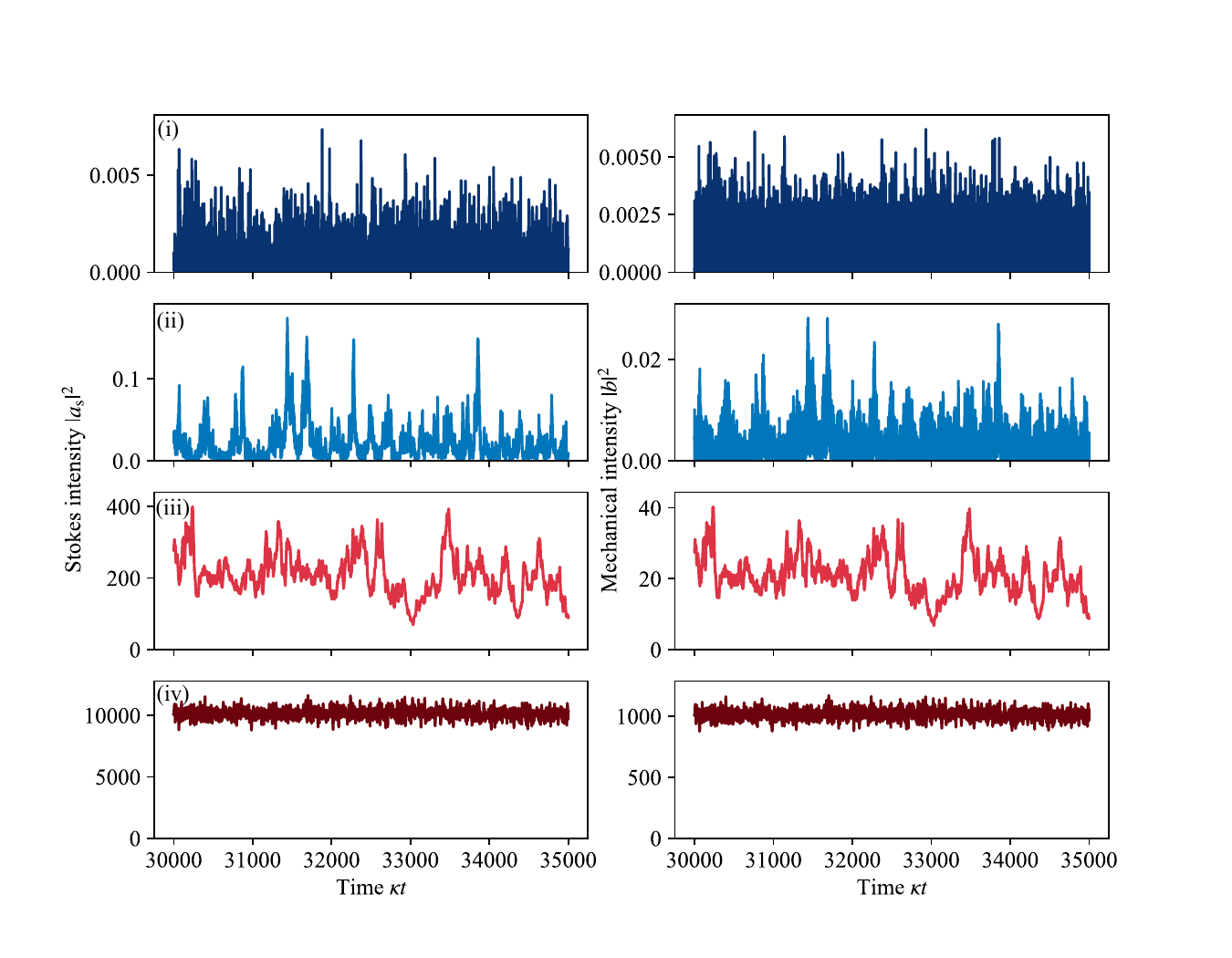}
    \caption{Simulated time series of Stokes $\left|a_{\text{S}}\right|^2$ and mechanical intensities $\left|b\right|^2$, the former of which is used to calculate the statistical average of $g^{(2)}(0)$. The simulations are for the same input powers as illustrated in the main text corresponding to the cases (i-iv).}
    \label{fig:SupFig1}
\end{figure}

The dynamics described by Eq.~(\ref{eqn:HLeqnsfull}) are numerically solved with the Euler method using Markovian input noise for both the thermal bath of the mechanics and the pump mode, the latter assumed to arise from frequency fluctuations of the pump laser. Due to thermal contributions in the Stokes scattering process dominating over vacuum fluctuations at room temperature, vacuum noise in the all modes is neglected from the numerical integration. The external input fields to the mechanical and pump modes at time $t$ are given by
\begin{align}
    \begin{split}
        b_{\text{in}}(t) &\sim \mathcal{N}(0,\Bar{n}) \ , \\
        a^{\text{(e)}}_{\text{in,P}}(t) &\sim \mathcal{N}(\alpha_{\text{in}},\Delta\alpha_{\text{in}}^2) \ ,
    \end{split}
\end{align}
where $\mathcal{N}(x,y)$ is a complex-valued normally-distributed random number of mean $x$ and variance $y$, $\alpha_{\text{in}}$ is the coherent amplitude of the input pump laser field, $\Delta\alpha_{\text{in}}^2$ is the variance of pump amplitude fluctuations and $\Bar{n}$ is the thermal occupation of the mechanical bath. The value of the noise is resampled in every integration time step and as the system can be treated classically and vacuum noise has been neglected, all operators are treated as C-numbers. The second-order coherence $g^{(2)}(0)$ is calculated from a long-time average after a period for which a steady state would have been reached. The simulation parameters are the same as the experimentally-determined parameters given in Table \ref{tab:ExperimentalParameters} and the system is initialised with zero field in each mode. Numerical solutions to Eq.~(\ref{eqn:HLeqnsfull}) at four different input powers and the calculated second-order coherence are displayed in Fig. 2 of the main text and example time traces at the same input powers for both the Stokes and the acoustic modes are illustrated in Fig. \ref{fig:SupFig1}.

% The noise value obtained corresponds to a fluctuation in intracavity photon number of 1 part in 50 at the onset of lasing.

%\begin{figure}
%    \centering
%    \includegraphics[width=\textwidth]{SupFigure1completev8.pdf}
%    \caption{Simulated $g^{(2)}(0)$ with input pump power normalised to the lasing threshold (above) and the time series of $\left|a_{\text{S}}\right|^2$ from which the statistical average is calculated (below). Blue circles in the above plot correspond to experimental data points. The dashed lines (i-iv) in the upper plot correspond to the labelled time series in the plot below. (i) Far below threshold (dark blue), the Stokes field displays thermal intensity fluctuations with very little coherent amplitude. (ii) Within the region of instability (light blue), large spikes in intensity are visible as the Stokes mode begins to ``flicker'', displaying super-thermal statistics. (iii) Below the lasing threshold (light red), the flickering persists but is now on top of a coherent intensity, resulting in a second-order-coherence between thermal and Poissonian statistics. (iv) Finally, far above threshold (dark red), intensity fluctuations are small relative to the strong coherent intensity and the light displays Poissonian statistics.}
%    \label{fig:SupFig1}
%\end{figure}

\section{Pump mode fluctuations}
In this section we will consider how frequency noise in the input pump field results in intracavity intensity pump noise. We first consider an optical resonance with a Lorentzian intensity response function $I$ given by
\begin{equation}
    I(\delta_\text{L}) = \frac{2\kappa}{(\kappa^2 + \delta_\text{L}^2)} \ ,
\end{equation}
where $\delta_\text{L}$ is the detuning from its centre frequency and $2\kappa/2\pi$ is its full-width-at-half-maximum. We are interested in the change in intracavity-intensity as the detuning is perturbed around the value $\delta_{\text{L}}$ describing the locking point of the laser. Perturbing around $\delta_{\text{L}}$ as $\delta_\text{L} \rightarrow \delta_{\text{L}}\pm\delta$ and keeping terms to first order, we obtain
\begin{align}
    I(\delta_{\text{L}}\pm\delta) &= \frac{2\kappa}{\kappa^2 + \delta_{\text{L}}^2(1\pm\delta/\delta_{\text{L}})^2} \notag \\
    &= \frac{2}{\kappa}\frac{1}{1+(\delta_{\text{L}}/\kappa)^2(1\pm\delta/\delta_{\text{L}})^2} \notag \\
    &\simeq \frac{2}{\kappa} \left(1-\left(\frac{\delta_{\text{L}}}{\kappa}\right)^2\left(1\pm\frac{\delta}{\delta_{\text{L}}}\right)^2\right) \notag \\
    &\simeq \frac{2}{\kappa}\left(1-\left(\frac{\delta_{\text{L}}}{\kappa}\right)^2\pm2\left(\frac{\delta_{\text{L}}}{\kappa}\right)^2\frac{\delta}{\delta_{\text{L}}}\right)
\end{align}
where we have assumed, as in the experimental set-up, that $\delta_{\text{L}}/\kappa<1$ in order to go from the second to the third line. For a frequency variation of $\delta$, this would therefore result in an intensity variation of 
\begin{equation}
    \frac{\Delta I}{I} = \frac{\left|I(\delta_{\text{L}}+\delta)-I(\delta_{\text{L}}-\delta)\right|}{I(\delta_{\text{L}})} = \frac{4\delta_{\text{L}}}{\kappa^2} \delta \ .
\end{equation}
For the intensity fluctuations used in the numerical simulations ($\Delta I/I = 1/50
$), this corresponds to an intensity linewidth of $2\delta/2\pi=15$~kHz for a pump laser detuned 200~kHz from a 1.1~MHz optical resonance.

\section{Experimental Parameters}

\label{sec:ExperimentalParameters}

\begin{table}[h]
    \centering
    \caption{Table of experimental and system parameters.}
    \begin{tabular}{|c|c|}
        \hline
         \textbf{Parameter}& \textbf{Value}\\
         \hline
         Resonator diameter, $D_{\text{res}}$ & $280\,$ \textmu m\\
         Pump wavelength, $\lambda_\text{P}$ & 1550 nm\\
         Mechanical frequency, $\omega_{\text{M}}/2\pi $ & 11.0 GHz \\
         Pump linewidth, $2\kappa_{\text{P}}/2\pi$ & 1.1 MHz \\
         Stokes linewidth, $2\kappa_{\text{S}}/2\pi$ & 3.2 MHz \\
         Acoustic linewidth, $2\gamma/2\pi$ & 27.0 MHz \\
         Pump detuning, $\delta_{\text{L}}/2\pi$ & 200 kHz \\
         Input pump power, $P$ & $\sim40-100\,$ \textmu W \\
         Pump-enhanced coupling rate, $G/2\pi$ & $\sim2.44 - 4.65$ MHz \\
         Three-wave-mixing coupling rate, $g_0/2\pi$ &  309 Hz \\
         Pump intracavity occupation, $n_{\text{cav}}$ & $\sim10^8$ photons \\
         Sample temperature, $T$ & 295 K \\
         Mechanical bath occupation, $\Bar{n}$ & 558 \\
         Taper efficiency, $\eta_{\text{taper}}$ & 0.80 \\ 
         Filter efficiency, $\eta_{\text{filter}}$ & 0.15 \\
         SNSPD quantum efficiency, $\eta_{\text{SNSPD}}$ & 0.15 \\
         SNSPD countrate, $R_{\text{count}}$ & up to $\sim10^6$ s\textsuperscript{-1}\\
         \hline
    \end{tabular}
    \label{tab:ExperimentalParameters}
\end{table}
\end{document}